\documentclass[a4paper,11pt]{article}
\pdfoutput=1 % if your are submitting a pdflatex (i.e. if you have
\usepackage{jcappub} % for details on the use of the package, please
\usepackage[T1]{fontenc} % if needed

\usepackage{fancyhdr}
\usepackage{amssymb}
\usepackage{tikz}
\usepackage{graphicx}
\usepackage{multirow}
\usepackage{ulem}

\newcommand{\eq}[1]{Eq.~(\ref{#1})}
\newcommand{\eqs}[2]{Eqs.~(\ref{#1}) and (\ref{#2})}

\newcommand{\GeV}{\mathinner{\mathrm{GeV}}}

\def\bea{\begin{eqnarray}}
\def\eea{\end{eqnarray}}
\def\beq{\begin{equation}}
\def\eeq{\end{equation}}

\def\l{\left}
\def\r{\right}

\title{\boldmath New- vs. chaotic- inflations}

\author{Gabriela Barenboim}
\author{and Wan-Il Park}
\affiliation{Departament de F\'isica Te\`orica and IFIC, Universitat de Val\`encia-CSIC, \\ E-46100, Burjassot, Spain}
\emailAdd{Gabriela.Barenboim@uv.es}
\emailAdd{Wanil.Park@uv.es}

\abstract{We show that "spiralized" models of new-inflation can be experimentally identified mostly by their positive spectral running in direct contrast with most chaotic-inflation models which have negative runnings typically in the range of $\mathcal{O}(10^{-4}-10^{-3})$.
}

\begin{document}

\begin{flushright}
FTUV-15-03-17 \\
IFIC-15-31
\end{flushright}

\maketitle
\flushbottom

\section{Introduction}

In modern (or standard) cosmology, the idea of inflation \cite{Guth:1980zm} provides the most compelling solution to various problems  (e.g., flatness, horizon, and monopole problems) of old (big-bang) cosmology \cite{Guth:1980zm,Guth:1979bh}.
Also, the classicalized quantum fluctuations of the inflaton are regarded as the most plausible seed for the density perturbations in the present universe \cite{Mukhanov:1981xt,Mukhanov:1982nu}.
In this regard, the concept of inflation is now a kind of paradigm of modern cosmology.
In order to match the observations of our universe, the last primordial inflation should have at least about $50 \sim 60$ $e$-foldings, depending on the specifics on the thermal history of the Universe after that stage. 
Also, if its quantum fluctuations are responsible for the density perturbations at the present universe, the inflaton needs to roll down an almost flat potential which should also provide a smooth end of inflation.
There are numerous models of inflation, all fullfilling these requirements, but they are either theoretically unappealing or quite degenerate among themselves exhibiting only minor differences, well beyond the expected experimental resolution in the intermediate and near future. A situation that leaves us with little hope to get a hint on the shape of inflaton potential for quite some time.

Generically, inflation models can be categorized into two groups: \textit{small-field} and \textit{large-field} inflation.
In the former group, inflation takes place in a sub-Planckian regime of field space.
In the latter case, inflaton evolves over Planck scale.
In view of effective field theory, which most of inflation models belongs to, small-field inflations are the most sensible.
However, typically there are issues about the flatness of the inflaton potential (the so-called $\eta$-problem), fine tuning, or flat inconsistency with observations in the most simple models of both categories that leaves the category of single small field models quasi empty.
In recent years however, there have been interesting ideas on compactifying the trans-Planckian trajectory of the inflaton into a sub-Planckian regime of a two-dimensional field space by winding the trajectory \cite{Silverstein:2008sg,McAllister:2008hb,Berg:2009tg,McDonald:2014oza,McDonald:2014nqa,Li:2014vpa,Carone:2014cta,Barenboim:2014vea,Barenboim:2015zka,Li:2015mwa} which has put small field inflation back into the inflationary model building game.
These potentials are free from the $\eta$-problem although some amount of tuning seems still unavoidable.
In those scenarios, even though inflaton dynamics takes place in a two-dimensional field space, it is effectively the same as the case of single field inflation in the sense that the trajectory does not have any peculiar change during inflation.
From now on, we call these two-dimensional extensions of single-field inflation as \textit{spiralilzed} inflation. 
%although we used this term for a specific model in Ref.~\cite{Barenboim:2014vea}.
In terms of the canonically normalized inflaton field, the various type of potentials in spiralized inflation can have peculiar (e.g. fractional) power dependences on the inflaton field.

In the circumstance of having spiralized inflation scenarios, with fields always in the sub Planckian regime, one may wonder whether these models can be degenerate with models among their class or even with models in the large single-field one, rendering the distinction between small and large field meaningless.  Therefore the relevance of spiralized models largely depends on a positive answer to the question of whether discriminating models, in particular, models of new-inflation-type (spiralized models) from ones of chaotic-inflation-type is possible.     
If so, such a distinction will not only remove a half of the parameter space in inflation-model-building but also shed some
ligth on the mechanism of inflation.
%Indeed, as we will show, such a distinction  turns out to be possible at the level of the running of the spectral index even when there is a degeneracy at the level of the three leading observables of inflation (power spectrum, spectral index, and tensor-to-scalar ratio).
In fact, in this work, we show that spiralized new-inflation models can be distinguished from various chaotic-inflation models at the level of the running of the spectral index of  density perturbations even if there is a degeneracy of the three leading observables of inflation (power spectrum, spectral index, and tensor-to-scalar ratio).

This paper is organized as follows.
In Section~\ref{sec:SI}, we provide a general phenomenological description of spiralized inflation.
In Section~\ref{sec:num-anal}, we show the differences of observables among various selected inflation models as a result of our numerical analysis.
In Section~\ref{sec:validity}, the validity of the single-field description of spiralized inflation is discussed.
In Section~\ref{sec:dis-con}, conclusions will be drawn.
Collections of formulas for the inflationary observables in terms of slow-roll parameters of single field inflation, and formulas of slow-roll parameters of selected models are provided in Appendix~\ref{app:obss} and~\ref{app:slow-roll-para}, respectively.

\section{Spiralized inflation}
\label{sec:SI}
Spiralized inflation models can be described by the potential,
\beq \label{V-SIs}
V = V_\phi + V_{\rm M}
\eeq
where $V_\phi$ is a function of $\phi$ only and 
\beq \label{V-M}
V_{\rm M} = \Lambda^4 \l[ 1 - \sin (\phi^n/M^n - \theta) \r]
\eeq
where $\phi$ and $\theta$ are regarded as the radial and angular degrees of freedom of a complex field, and $n \in \mathbb{N}$.
We assume $V_{\rm M}/V_\phi \ll 1$ at least during inflation, and consider only $n=1,2$ cases for simplicity.
The inflaton is expected to trace closely the minimum of the spiraling valley of the potential $V$.
In this case, $\partial V/\partial \phi=0$ gives
\beq \label{dVdphi-0}
\frac{\partial V_\phi}{\partial \phi} = \frac{n}{M} \l( \frac{\phi}{M} \r)^{n-1} \Lambda^4 \cos \theta_\phi \equiv \frac{f(\phi)}{\phi} \Lambda^4 \cos \theta_\phi
\eeq
where $\theta_\phi \equiv \phi^n/M^n - \theta$, and $f(\phi) \equiv n \l( \phi / M \r)^n$.
It leads to
\beq \label{dphi-dtheta-0}
\l[ \frac{\partial^2 V_\phi}{\partial \phi^2} - \frac{n-1}{\phi} \frac{\partial V_\phi}{\partial \phi} + f^2 \frac{\Lambda^4}{\phi^2} \sin \theta_\phi \r] d\phi 
= \l[ f \frac{\Lambda^4}{\phi} \sin \theta_\phi \r] d \theta
\eeq 
When the inflaton is trapped in the spiraling trench (i.e. $\phi \gtrsim M$ (or $f \gtrsim 1$)), the curvature along $\phi$ is dominated by the contribution from $V_{\rm M}$ in \eq{V-SIs}.
In this case, the last term in the left-hand bracket of \eq{dphi-dtheta-0} dominates the other terms, resulting in
\beq \label{dphi-dtheta}
d \phi \approx \phi d \theta/f
\eeq
which defines inflaton's trajectory. 

In the basis of ($\phi, \theta$), when the field configuration is constrained to follow a specific trajectory such that $\phi$ and $\theta$ are dependent on each other, an infinitesimal displacement along the trajectory is defined as
\beq \label{dI}
dI \equiv \l[1 + \l( \frac{\phi d \theta}{d \phi} \r)^2 \r]^{1/2} d\phi = \l[1 + \l( \frac{d \phi}{\phi d \theta} \r)^2 \r]^{1/2} \phi d\theta
\eeq
The unit vectors along the trajectory ($I$) and the orthogonal direction can be written as  
\beq
\boldsymbol{e}_I^T = (c_\phi, \ c_\theta), \quad \boldsymbol{e}_\perp^T = (c_\theta, \ - c_\phi) 
\eeq
Then, the directional derivative along inflaton is given by
\beq
\frac{d}{d I} = \boldsymbol{e}_I \cdot \nabla = c_\phi \frac{\partial}{\partial \phi} + c_\theta \frac{\partial}{\phi \partial \theta}
\eeq
where
\bea
c_\phi &\equiv& \frac{\partial \phi}{\partial I} = \frac{d \phi/d \theta}{\sqrt{\phi^2 + \l( d \phi/d \theta \r)^2}}
\\
c_\theta &\equiv& \frac{\phi \partial \theta}{\partial I} = \frac{\phi}{\sqrt{\phi^2 + \l( d \phi/d \theta \r)^2}} 
\eea
The slope along the direction is 
\beq
\frac{dV}{dI} = c_\phi \frac{\partial V}{\partial \phi} + c_\theta \frac{\partial V}{\phi d \theta} 
\eeq
and the mass is obtained as  
\beq
\frac{d^2 V}{d I^2} = c_\phi^2 \mathbb{M}_{\phi \phi}^2 + 2 c_\phi c_\theta \mathbb{M}_{\phi \theta}^2 + c_\theta^2 \mathbb{M}_{\theta \theta}^2
\eeq
where the mass matrix elements are found to be
\bea
\mathbb{M}^2_{\phi \phi} &=& \frac{\partial^2 V}{\partial \phi^2} + \frac{\partial \ln c_\phi}{\partial \phi} \frac{\partial V}{\partial \phi} 
\\
\mathbb{M}^2_{\phi \theta} &=& \frac{\partial^2 V}{\phi \partial \theta \partial \phi} - \frac{1}{2} \l( 1 - \frac{\partial \ln c_\theta}{\partial \ln \phi} \r) \frac{\partial V}{\phi^2 \partial \theta} + \frac{\partial \ln c_\phi}{\partial \theta} \frac{\partial V}{\phi \partial \phi}
\\
\mathbb{M}^2_{\theta \theta} &=& \frac{\partial^2 V}{\phi^2 \partial \theta^2} + \frac{\partial \ln c_\theta}{\partial \theta} \frac{\partial V}{\phi^2 \partial \theta}
\eea

Along the spiraling inflaton direction following \eq{dphi-dtheta} and being expected to satisfy $\partial V/\partial \phi = 0$ with a good accuracy,  $c_\phi$ and $c_\theta$ can be regarded as functions of $\phi$ only, and one finds 
\bea
\frac{\partial^2 V}{\partial \phi^2} &=& \frac{\partial^2 V_\phi}{\partial \phi^2} - \frac{n-1}{\phi} \frac{\partial V_\phi}{\partial \phi} + f^2 \frac{\Lambda^4}{\phi^2} \sin \theta_\phi 
\\
\frac{\partial^2 V}{\phi \partial \theta \partial \phi} &=& - f \frac{\Lambda^4}{\phi^2} \sin \theta_\phi
\\
\frac{\partial V}{\phi^2 \partial \theta} &=& \frac{\Lambda^4}{\phi^2} \cos \theta_\phi 
\\
\frac{\partial^2 V}{\phi^2 \partial \theta^2} &=& \frac{\Lambda^4}{\phi^2} \sin \theta_\phi
\eea
leading to
\bea \label{dVdI}
\frac{dV}{dI} &=& \gamma V_\phi^{'}
\\ \label{d2VdI2}
\frac{d^2 V}{d I^2} &=& \gamma^2 \l[ V_\phi^{''} - n \frac{V_\phi^{'}}{\phi} + n \gamma^2 \frac{V_\phi^{'}}{\phi} \r]
\\ \label{d3VdI3}
\frac{d^3 V}{d I^3} &=& \gamma^3 \l\{ V_\phi^{'''} - 3n \frac{V_\phi^{''}}{\phi} + n \l( 2 n + 1 \r) \frac{V_\phi^{'}}{\phi^2} + n \gamma^2 \l[ 3 \frac{V_\phi^{''}}{\phi} -  \l( 6n + 1 \r) \frac{V_\phi^{'}}{\phi^2} \r] + 4 n^2 \gamma^4 \frac{V_\phi^{'}}{\phi^2} \r\}
\\ \label{d4VdI4}
\frac{d^4 V}{d I^4} &=& \gamma^4 \l\{ V_\phi^{''''} - 6n \frac{V_\phi^{'''}}{\phi} + n (11 n + 4) \frac{V_\phi^{''}}{\phi^2} - n (2 n + 1)(3n+2) \frac{V_\phi^{'}}{\phi^3} \r.
\nonumber \\
&& \phantom{\gamma^4 \l\{ \ \r.} \l. + n \gamma^2 \l[ 6 \frac{V_\phi^{'''}}{\phi} - (30n+4) \frac{V_\phi^{''}}{\phi^2} + (36n^2+20n+2) \frac{V_\phi^{'}}{\phi^3} \r] \r.
\nonumber \\
&& \phantom{\gamma^4 \l\{ V_\phi^{''''} - 6n \frac{V_\phi^{'''}}{\phi} \ \r.} \l. + n^2 \gamma^4 \l[ 19 \frac{V_\phi^{''}}{\phi^2} - (58n+13) \frac{V_\phi^{'}}{\phi^3} \r] + 28 n^3 \gamma^6 \frac{V_\phi^{'}}{\phi^3} \r\}
\eea
where $\gamma \equiv c_\theta/f \ll 1$, `$'$' denotes derivative with respect to $\phi$, and $d \ln c_\theta/ d \ln \phi = n \gamma^2$ was used.
In these derivatives of $V$ with respect to $I$ we kept all higher order terms of $\gamma$, since the leading order contributions can be cancelled out, depending on $V_\phi$ and $n$.

Alternatively, in the region where \eq{dphi-dtheta} is valid, the inflaton can be expressed as
\beq \label{inflaton}
I = \int \frac{f}{c_\theta} d \phi \approx \int f d\phi = \frac{n}{n+1} \l( \frac{\phi}{M} \r)^{n+1} M
\eeq
where we assumed $f \gg 1$ which was justified with numerical tests.
For $\phi \gg M$, if $\phi$ is away from the end point of inflation and $\theta_\phi$ is nearly constant for several $e$-foldings associated to the observed CMB scales, one may ignore the contribution of $V_M$ in the potential $V$ of \eq{V-SIs} as long as $V_M \ll V_\phi$.
In this case, \eq{inflaton} allows a simple single-field description of spiralized inflation.
Note however that such a description implies setting $d \ln c_\theta/ d \ln \phi=0$ which is not problematic in many cases, but can lead to a wrong result in some cases (e.g., `Spiral inflation' with $n=1$).
Also, for a tachyonic $V_\phi$, as the inflaton evolves close to the end of inflation, $|\partial V_\phi/\partial \phi|$ becomes larger and the last term in the left-hand side bracket of \eq{dphi-dtheta-0} can become subdominant, depending on $M$ and $\Lambda$.
In such a case, \eq{dphi-dtheta} does not hold any more, and it becomes non-trivial to find out a simple single-field description for the inflation along the inflaton's trajectory.
Hence, we do not take this approach, but will use Eqs.~(\ref{dVdI})-(\ref{d4VdI4}) in order to obtain analytic expressions for the  slow-roll parameters.
The observables of spiralized inflation in terms of the slow-roll parameters can be found in the same way as in the single-field case (see Appendix~\ref{app:obss}). 
Explicit expressions of slow-roll parameters for several selected models of our interest can be found in Appendix~\ref{app:slow-roll-para}.

As a remark, a general feature of spiralized inflation is that the $n$-th derivative of potential with respect to the canonical inflaton field is suppressed by $f^n$ relative to the case without spiral motion (i.e. the case of $V_{\rm M}=0$).
This results in suppressions of slow-roll parameters relative to the case of usual single-field inflation, allowing effective large single-field slow-roll inflation in sub-Planckian field space.

\section{Numerical analysis}
\label{sec:num-anal}

In this section, a broad choice of models of large-field inflation and spiralized inflation is presented and analyzed numerically to see the possibility of discriminating among different models.
For slow-roll parameters, formulas collected in Appendix~\ref{app:slow-roll-para} were used.

\subsection{Models}
We choose following models for comparison:
\begin{itemize}
\item Hilltop inflation (HI) \cite{Boubekeur:2005zm}:
\beq
V=V_\phi = V_0 \l[ 1 - \l( \frac{\phi}{\mu} \r)^4 \r] + \dots
\eeq
\item $R^2$-inflation (R2I) \cite{Starobinsky:1980te}:
\beq \label{R2I}
V=V_\phi= V_0 \l( 1 - e^{-\phi/\mu} \r)^2
\eeq
\item Natural inflation (NI) \cite{Freese:1990rb}:
\beq \label{NI}
V = V_\phi = V_0 \l[ 1 + \cos(\phi/\mu) \r]
\eeq
\item Spiral chaotic inflation 1 (SCI1) \cite{Berg:2009tg}:
\beq
V = V_0 \l( \frac{\phi}{\mu} \r)^2 + V_{\rm M}
\eeq
\item Spiral chaotic inflation 2 (SCI2) \cite{McDonald:2014nqa}:
\beq
V = V_0 \l( \frac{\phi}{\mu} \r)^4 + V_{\rm M}
\eeq
\item Spiral inflation (SI) \cite{Barenboim:2014vea}:
\beq
V = V_0 \l[ \l( \frac{\phi}{\phi_0} \r)^2 - 1 \r]^2 + V_{\rm M}
\eeq
\item Spiral Coleman-Weinberg inflation (SCWI) \cite{Barenboim:2015zka}:
\beq
V = V_0 \l\{ 1 + 4 \l( \frac{\phi}{\phi_0} \r)^4 \l[ \ln \l( \frac{\phi}{\phi_0} \r) - \frac{1}{4} \r] \r\} + V_{\rm M}
\eeq
\end{itemize}
Some of well-known large-field models have been excluded from this selection, since they seem unlikely to be consistent with recent data from Planck satellite mission \cite{Ade:2015lrj}.

\subsection{Distribution of models on ($n_s,r$)- and ($\alpha_\mathcal{R},\alpha_\mathcal{R}'$)-planes}
Many simple models of single-field inflation can be distinguished by its spectral index ($n_s$) and tensor-to-scalar ratio ($r$).
However, there can be degeneracy among some models at the $n_s$ and $r$ level.
In this case, the next thing we should see is the running of spectral index ($\alpha_\mathcal{R}$), or one may have to go even further (e.g., to the running of the running ($\alpha_\mathcal{R}' \equiv d \alpha_\mathcal{R}/d \ln k$)).
In terms of $n_s$ and $r$, one finds
\bea \label{alphaR}
\alpha_{\mathcal{R}} &=& - \frac{1}{2} r \l( 1-n_s-\frac{3}{16}r \r) - 2 \xi^2
\\ \label{dalphaR}
\alpha_\mathcal{R}' &=& -\frac{1}{2} r \l[ \l( 1 - n_s \r)^2 - \frac{3}{64} r^2 \r] - \l( 1 - n_s + \frac{9}{8} r \r) \xi^2 + 2 \sigma^3
\eea
Note that in \eq{alphaR} the first term of the right-hand side is always negative for $r<0.1$ and $0.95 \lesssim n_s \lesssim 0.98$ \cite{Ade:2015lrj}.
Hence, it may be tempting to use the sign of $\xi^2$ as a  discriminator between models of inflation at the level of $\alpha_\mathcal{R}$ although only an analysis about the magnitude of $\xi^2$ will be able to tell, when it is negative, whether it is a good discriminator.
$\sigma^3$ in \eq{dalphaR} may also play a role similar to $\xi^2$ in \eq{alphaR} but in combination with $\xi^2$.
It is thus instructive to categorize the  generic behaviors of $\xi^2$ and $\sigma^3$ in several prototype simple potentials.
It is straightforward to see that for
\begin{itemize}
\item Monomial large-field models: $|\xi^2| \lesssim \mathcal{O}(\epsilon^2)$ and $|\sigma^3| \lesssim \mathcal{O}(\epsilon^3)$, leading to $\alpha_\mathcal{R} < 0$ and $d \alpha_\mathcal{R}/d \ln k < 0$.
\item Concave chaotic-inflation models: $dV/dI>0$ and $d^3V/dI^3>0$, leading to $\xi^2>0$ and hence $\alpha_\mathcal{R}<0$.
\item Concave new-inflation models: $dV/dI<0$ and $d^3V/dI^3>0$, leading to $\xi^2<0$. Hence there is possibility of $\alpha_\mathcal{R}>0$.
\end{itemize}
As can be seen from \eq{alphaR}, since $\alpha_\mathcal{R}$ is an observable, $\xi^2$ should be physical quantity, too.
Hence, once the sign of $dV/dI$ is fixed, that of $d^3V/dI^2$ is fixed too, irrespective of possible field redefinitions.  
Note that only concave new-inflation models display the  possibility of having a positive spectral running. 

In Table~\ref{tab:V3rd}, we show the patterns of $\xi^2$ and $\sigma^3$ for the various models of interest.
\begin{table}[h!]
\begin{center}
\begin{tabular}{|c|c||c|c|c|}
\hline
& Model & $V$ & Sign($\xi^2$) & Sign($\sigma^3$)\\
\hline \hline
%\multicolumn{1}{ |c }{\multirow{4}{*}{Single-field}} & \multicolumn{1}{ |c|| }{CI1} & $\frac{1}{2} m^2 \phi^2$ & $0$ & $0$\\
%\cline{2-5}
%\multicolumn{1}{ |c }{} & \multicolumn{1}{ |c|| }{CI2} & $\frac{1}{4} \lambda \phi^4$ & $+$ & $+$\\
%\cline{2-5}
\multicolumn{1}{ |c }{\multirow{3}{*}{Single-field}} & \multicolumn{1}{ |c|| }{HI} & $V_0 \l[ 1 - \l( \frac{\phi}{\mu} \r)^4 \r] + \dots$ & $+$ & $-$ \\
\cline{2-5}
\multicolumn{1}{ |c }{} & \multicolumn{1}{ |c|| }{R2I} & $V_0 \l( 1 - e^{-\phi/\mu} \r)^2$ & $+$ & $-$ \\
\cline{2-5}
\multicolumn{1}{ |c }{} & \multicolumn{1}{ |c|| }{NI} & $V_0 \l[ 1 + \cos(\phi/\mu) \r]$ & $-$ & $-$ \\
\hline
\multicolumn{1}{ |c }{\multirow{4}{*}{Two-field}} & \multicolumn{1}{ |c|| }{SCI1} & $V_0 (\phi/\mu)^2 + V_{\rm M}$& $-(+)$ & $+(-)$ \\
\cline{2-5}
\multicolumn{1}{ |c }{} & \multicolumn{1}{ |c|| }{SCI2} & $V_0 (\phi/\mu)^4 + V_{\rm M}$& $+(-)$ & $-(+)$ \\
\cline{2-5}
\multicolumn{1}{ |c }{} & \multicolumn{1}{ |c|| }{SI} & $V_0 \l[ \l( \frac{\phi}{\phi_0} \r)^2 - 1 \r]^2 + V_{\rm M}$ & $-$ & $-$\\
\cline{2-5}
\multicolumn{1}{ |c }{} & \multicolumn{1}{ |c|| }{SCWI} & $V_0 \l\{1 + 4 \l( \frac{\phi}{\phi_0} \r)^4 \l[ \ln \l( \frac{\phi}{\phi_0} \r) - \frac{1}{4} \r] \r\} + V_{\rm M} $ & $-(-)$ & $-(-)$\\
\hline
\end{tabular}
\end{center}
\caption{Patterns of $\xi^2$ and $\sigma^3$ for sample models of inflation.
The sign in parenthesis is for $n=2$ case.
In R2I, $\mu=\sqrt{3/2} M_{\rm P}$.
In NI, the sign of $\sigma^3$ depends on $\mu$, but as $\mu$ becomes much larger than $M_{\rm P}$, $\cos(\phi/\mu)$ becomes negative leading to a negative $\sigma^3$. 
In SI and SCWI, we took $\phi_0=M_{\rm P}$ and $M_{\rm GUT}$, respectively.}
\label{tab:V3rd}
\end{table}%
%
%--------------------------
\begin{figure}[h!]
\begin{center}
\includegraphics[width=0.49\textwidth]{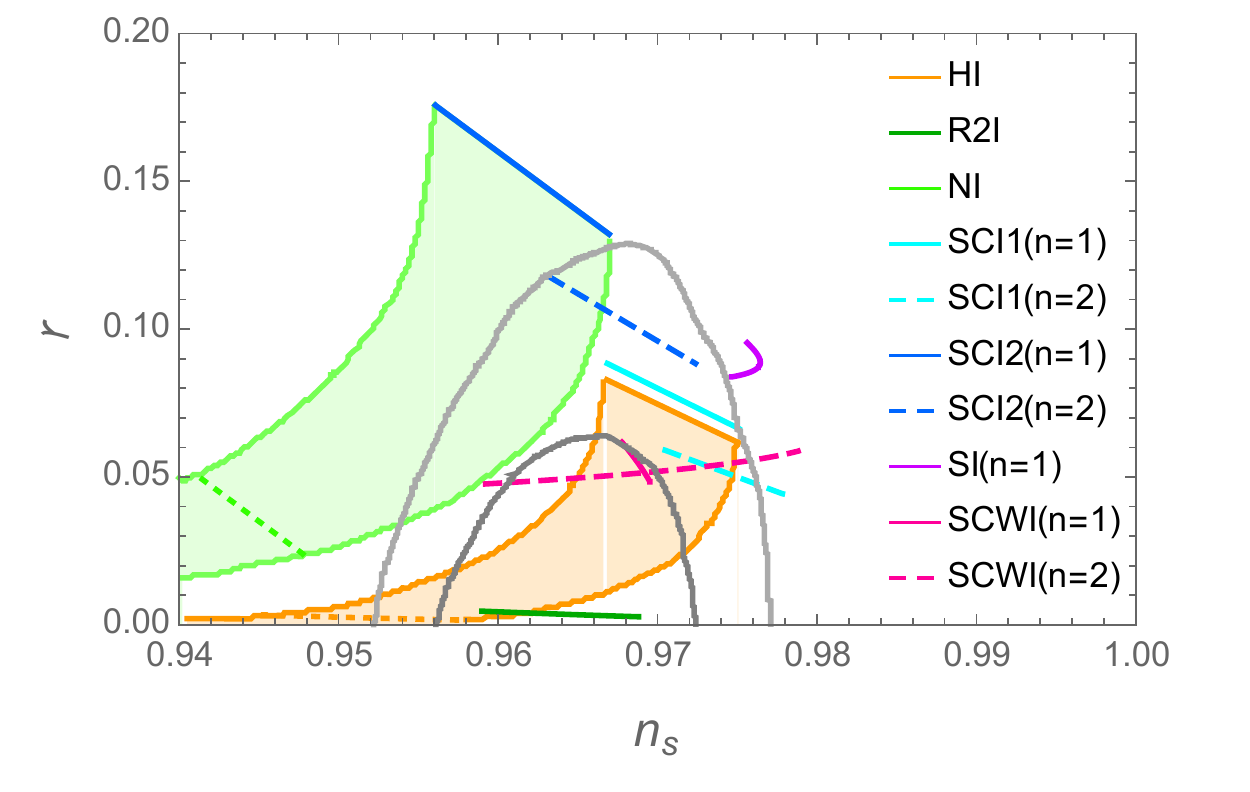}
\includegraphics[width=0.49\textwidth]{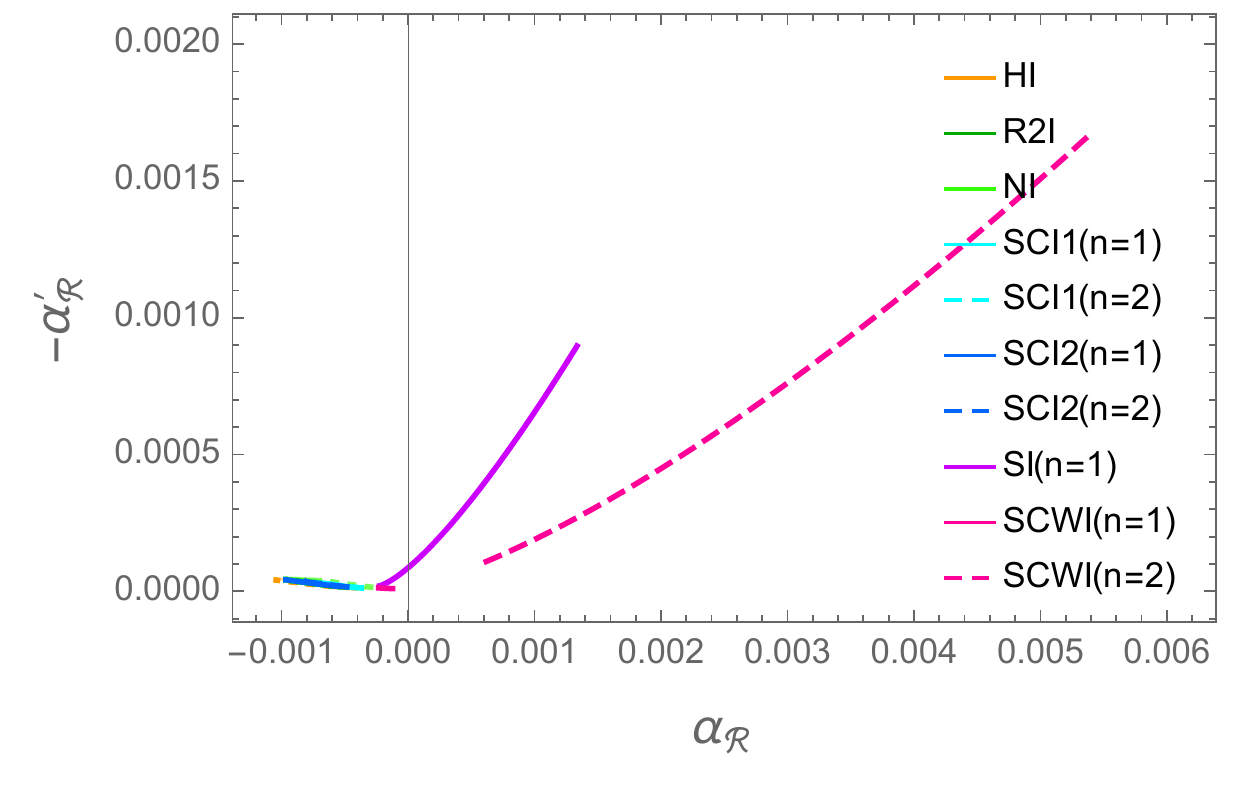}
\includegraphics[width=0.49\textwidth]{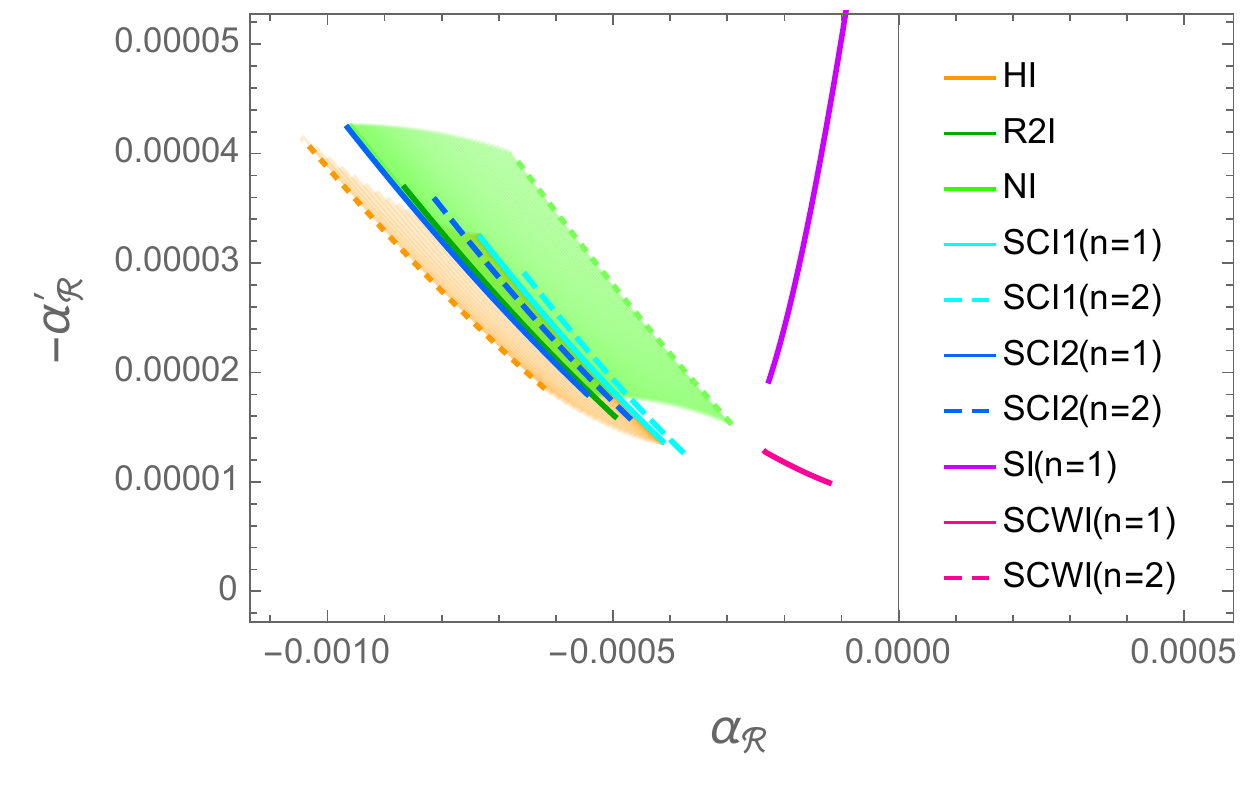}
\caption{Observables. 
\textit{Top-Left}: Dark-gray and gray lines are respectively $1\sigma$ and $2\sigma$ CLs of Planck data \cite{Ade:2015lrj}. 
A colored straight line corresponds to $N_e=45-60$ from right to left end except the cases of `SI' and `SCWI'.
For spiralized new-inflation models, each line covers $\Delta N_e = 10$ starting from $N_e = 39.5, 45$ for `SI($n=1$)' and `SCWI($n=1,2$)', respectively. 
Green and orange shade-regions correspond respectively to $\mu \leq 250$ and $\mu \leq 300$ from right.
Green and orange dotted lines correspond to $\mu=4.65$ and $7$, respectively.
For spiralized inflation models, we took the parameter $M$ appearing in the modulating potential $V_{\rm M}$ as shown in Table~\ref{tab:M-value}. 
Note that ``SI'' has only $n=1$ case, since $n=2$ is phenomenologically ruled out.
\textit{Top-Right}: Same color scheme and parameter set as the left panel.
\textit{Bottom}: Same as top-right panel in a different scale. 
Green and orange regions cover $\mu = 4.65-250$ and $7-300$ with dotted lines corresponding to $\mu=4.65$ and $7$, repectively.
Some right part of the orange region is overlapped by the green region.
Large parts of solid blue and dark green lines are nearly overlapped.
}
\label{fig:obss}
\end{center}
\end{figure}
%---------------------------
%
Now it is easy to see that Hilltop- and $R^2$-inflation are expected to have negative $\alpha_\mathcal{R}$.
On the other hand, it may be possible for SI and SCWI to have positive $\alpha_\mathcal{R}$ since their $\xi^2$ is negative.  
That is, spiral new-inflations may be distinguished from the others.
Since the signs of $\xi^2$ and $\sigma^3$ are only suggestive and the sign of $\alpha_\mathcal{R}$ depends on the specific form of potential $V_\phi$, a numerical analysis is required.
We calculated numerically inflationary observables up to $\alpha_\mathcal{R}'$, using formulas collected in Appendix A and B.
As a result of the analysis, we show the positions of models in ($n_s,r$)- and ($\alpha_\mathcal{R}, \alpha_\mathcal{R}'$)-planes in Fig.~\ref{fig:obss}.
In the top left panel of the figure, one see that most of models we have considered (which are a fair sample of what can be found in the literature) can be distinguished in ($n_s, r$)-plane, but there are still several models which may be difficult to be distinguished by $n_s$ and $r$.
It is interesting to see that in SCWI $r \sim 0.05$ even if we took $\phi_0 = M_{\rm GUT}$.
This is because the spiraling motion of inflaton extends $e$-foldings such that, for the last about $50$ $e$-foldings, inflaton had to be close to $\phi_0$, where $\epsilon$ is sizable \footnote{This result is different from Ref.~\cite{Barenboim:2015zka} because in this work the waterfall end point of inflation was pushed maximally toward $\phi_0$.
In this case, another inflation can take place successively with a large amount additional $e$-foldings, since inflaton would get trapped again.
Since inflation can end much earlier by making the modulating potential small, we put aside this issue in this work.}.   
In the top right and bottom panels, we notice that the degeneracy in ($n_s,r$)-plane is mostly broken at $\alpha_\mathcal{R}$.
The pattern of $\alpha_\mathcal{R}'$ in regard of $\xi^2$ and $\sigma^3$ is not so clear because of the fact that some models display a somewhat large $n_s$ and $r$.
However, notably, spiralized new inflation models can have quite large positive $\alpha_\mathcal{R}$ (depending on $n$) and large negative $\alpha_\mathcal{R}'$.
This behaviour can be seen from Eqs.~(\ref{SI-xi2}), (\ref{SI-sigma3}), (\ref{SCWI-xi2}) and (\ref{SCWI-sigma3}) with $\phi_* \sim \mathcal{O}(0.1) \phi_0$ as the field value giving observables consistent with data for the cosmological scales of interest. 
We have not considered some power-law potentials \cite{Linde:1983gd} which can be motivated from axion monodromy models \cite{Silverstein:2008sg,McAllister:2008hb}, but as commented already, its is straitforward to see that such potentials give $\alpha_\mathcal{R} \sim - \mathcal{O}(10^{-3})$ since $\epsilon = \mathcal{O}(10^{-3}-10^{-2})$ and $\xi^2 \sim \pm \mathcal{O}(\epsilon^2) (\textrm{or} \ 0)$.
Therefore, we see that spiral new-inflation can be clearly distinguished from all the other chaotic models of inflation at the level of $\alpha_\mathcal{R}$ at least.
%---------------
\begin{table}[h!]
\begin{center}
\begin{tabular}{|c||c|c|c|c|}
\hline
& SCI1 & SCI2 & SI & SCWI \\
\hline \hline
$n=1$ & $1.0 \times 10^{-3}$ & $2.0 \times 10^{-2}$ & $ 2.52 \times 10^{-2} $ & $ 2.0 \times 10^{-4}$ \\
\hline
$n=2$ & $ 2.0 \times 10^{-2}$ & $ 5.0 \times 10^{-2} $ & $ - $ & $ 1.48 \times 10^{-2}$ \\
\hline
\end{tabular}
\end{center}
\caption{The value of $M$ in $V_M$ of \eq{V-M} used for the numerical analysis of spiralized inflation models.
The unit is $M_{\rm P}=2.4 \times10^{18} \GeV$ except SCWI for which $M_{\rm GUT} = 2 \times 10^{16} \GeV$ was used.}
\label{tab:M-value}
\end{table}%
%----------------------

Future measurements of the power spectrum of cosmological weak lensing as the one that will be performed by the planned all-sky optical  EUCLID \cite{EUCLID}  might detect a running of the primordial spectral index at the required level ($\mathcal{O}(10^{-4})$), 
provided the uncertainties about the source redshift distribution and the underlying matter power spectrum are under control. 
The NASA SPHEREx mission \cite{SPHEREx}, a  proposed  all-sky  spectroscopic  survey, forectast a factor of 2 improvement over EUCLID perspective offers an ideal tool for discriminating models.

\section{The validity of the single field description of spiralized inflation}
\label{sec:validity}

The validity of the single field description of spiralized inflation can be guaranteed when the dynamics along the direction orthogonal to inflaton can be ignored and the turning rate of the inflaton trajectory is much smaller than unity.
The mass-square along the orthogonal direction is found to be 
\bea
\frac{d^2 V}{d \psi^2} 
&=& c_\theta^2 \mathbb{M}_{\phi \phi}^2 - 2 c_\phi c_\theta \mathbb{M}_{\phi \theta}^2 + c_\phi^2 \mathbb{M}_{\theta \theta}^2
\nonumber \\
&=& c_\theta^2 \l[ V_\phi^{''} + \l( f + \frac{1}{f} \r)^2 \frac{\Lambda^4}{\phi^2} \sin \theta_\phi  - f \l( n - 1 - \frac{1 - n \gamma^2}{f^2} \r) \frac{\Lambda^4}{\phi^2} \cos \theta \r]
\approx f^2 \frac{\Lambda^4}{\phi^2} \sin \theta_\phi
\eea
where we assumed $f\gg1$ and $\sin \theta_\phi \gtrsim \cos \theta_\phi$ which are valid for the cosmological scales of interest.
If the mass scale along $\psi$ is comparable to or larger than the expansion rate during inflation, the background motion and perturbations along $\psi$ are exponentially damped out within a few $e$-foldings.
This requires
\beq \label{m-orthgo-H}
\frac{d^2 V/d \psi^2}{3 H^2} \simeq f^2 \l( \frac{M_{\rm P}}{\phi_*} \r)^2 \frac{\Lambda^4}{V(\phi_*)} \sin \theta_\phi \gtrsim 1
\eeq
where $\phi_*$ is the field value when a cosmological scale of interest exits the horizon during inflation.
Note that the first slow-roll parameter $\epsilon$ can be expressed as 
\beq \label{ep-gen}
\epsilon = \frac{1}{2} \l| \frac{M_{\rm P}}{V} \frac{dV}{dI} \r|^2 = \frac{1}{2} \l( \frac{c_\theta}{f} \frac{M_{\rm P}}{\phi_*} \frac{\phi_* V_\phi^{'}}{V} \r)^2
\eeq
Also, for spirlized models we are considering, $\phi_* V_\phi^{'} \sim \phi_*^2 V_\phi^{''}$ within a factor of a few.
Hence, \eq{m-orthgo-H} together with \eqs{dVdphi-0}{ep-gen} can be interpreted as
\beq \label{m2-phi}
\l( \frac{V_\phi^{''}}{3 H^2} \r)^2 \gtrsim \sqrt{\frac{r}{8}} \l( \frac{M_{\rm P}}{\phi_*} \r) \cot \theta_\phi
\eeq
In \eq{m2-phi}, the left-hand side for a sub-Planckian excursion of $\phi$ is generically of $\mathcal{O}(1)$ or larger.
The right-hand side is typically smaller than or at most comparable to unity. 
Hence, the dynamics along the orthogonal direction can be safely ignored.

Also, following Ref.~\cite{Peterson:2010mv}, the turning rate is found to be
\beq
\frac{\eta_\perp}{v} \simeq - \frac{n \sqrt{2 \epsilon}}{f} \l( \frac{M_{\rm P}}{f \phi} \r)
\eeq
where $\eta_\perp$ and $v$ are respectively the field acceleration to the direction orthogonal to infaton and the field speed measured with respect to $e$-foldings instead of time.
For a simple power-law potential of $V_\phi$, $M_{\rm P}/f \phi \sim \sqrt{2 \epsilon} \lesssim \mathcal{O}(0.1)$ for cosmological scale of interest.
Even for SI and SCWI, we find that $M_{\rm P}/f^2 \phi \sim \mathcal{O}(10^{-2})$.
Hence, for $\gamma \lesssim \mathcal{O}(0.1)$ in our analysis, we find 
\beq
\frac{\eta_\perp}{v} \lesssim \mathcal{O}(10^{-3}) 
\eeq
Therefore, the single field description is a good approximation for spiralized inflation models, and non-Gaussianities are expected to be much smaller than unity.

\section{Discussions and conclusions}
\label{sec:dis-con}

In this paper, we studied the possibility of discriminating models of inflation by taking a look at the pattern of $\alpha_\mathcal{R}$ (the spectral running) and $\alpha_\mathcal{R}'$ (the running of the running) of the density perturbations originated from the quantum fluctuations of the inflaton field.
As sample models, several large-field inflation models (Hilltop- and $R^2$-inflation) including natural inflation and spiralized inflation models (spiral new- and chaotic-inflations) were considered.
As a result of our numerical analysis, we found that all the chaotic models selected (including natural inflation) have negative definite spectral runnings of $\mathcal{O}(10^{-4}-10^{-3})$, while spiral new-inflation models mostly have positive $\alpha_\mathcal{R}$s which can be as large as a few times $\mathcal{O}(10^{-3})$.
Also, spiral new-inflation models can have very large $|\alpha_\mathcal{R}'|$s a fact that allows easy discrimination of the models in future experiments, although they might be also discriminated from their spectral indices and tensor-to-scalar ratios.
Hence it will be easy to rule out either new-inflation-type model or chaotic-inflation-type ones in future observational experiments once the experimental uncertainties on $\alpha_\mathcal{R}$ go below $\mathcal{O}(10^{-3})$.

\appendix

\section{Observables in single-field inflation}
\label{app:obss}
The \textit{slow-roll inflation} is the simplest way of generating a nearly scale-invariant power spectrum of density perturbation via a slowly rolling single inflaton field ($I$).
In the slow-roll limit, the equation of motion of inflaton is approximated as
\beq
0 = \ddot{I} + 3 H \dot{I} + V' \approx 3 H \dot{I} + V'
\eeq
and inflation is characterized by slow-roll parameters defined as
\bea
\epsilon &\equiv& \frac{1}{2} \l( \frac{M_{\rm P} V'}{V} \r)^2
\\
\eta &\equiv& \frac{M_{\rm P}^2 V''}{V}
\\
\xi^2 &\equiv& \frac{M_{\rm P}^4 V' V'''}{V^2} 
\\
\sigma^3 &\equiv& \frac{M_{\rm P}^6 V'^2 V''''}{V^3} 
\eea
where derivatives denoted by `$\prime$'s are with respect to the inflaton field ($I$).
%, and ``SFI" and ``LFI" represents small-field and large-field inflation, respectively.
The $e$-folding number of a slow-roll inflation is given by
\beq
N_e = \int_{t}^{t_e} H dt \approx - \frac{1}{M_{\rm P}^2} \int_{I}^{I_e} \frac{V}{V'} dI %= - \frac{1}{M_{\rm P}} \int_\phi^{\phi_e} \frac{d\phi}{\sqrt{2 \epsilon(\phi)}}
\eeq
where the subscript `$e$' stands for the end of inflation.
As observables, the density power spectrum and its special index of a slow-roll inflation are given by
\bea
P_{\mathcal{R}} &\equiv& \l( \frac{H}{2 \pi} \r)^2 \l( \frac{\partial N_e}{\partial I} \r)^2 \approx \frac{1}{8 \pi^2} \frac{H^2}{\epsilon M_{\rm P}^2}
\\
n_s-1 &\equiv& \frac{d \ln P_{\mathcal{R}}}{d \ln k} \approx \l(2 \eta - 6 \epsilon \r) \l(1 + \epsilon \r)
\eea
For the tensor-mode, 
\bea
P_T &\equiv& \l( \frac{8}{M_{\rm P}^2} \r) \l( \frac{H}{2 \pi} \r)^2
\\
n_T &\equiv& \frac{d \ln P_T}{d \ln k} = - 2 \epsilon
\eea
The tensor-to-scalar ratio is given by
\beq
r \equiv \frac{P_T}{P_{\mathcal{R}}} = 16 \epsilon
\eeq
For a cosmological scale leaving the horizon at a given epoch, $d \ln k = d \ln (a H)$ leading to
\beq
\frac{d \ln k}{d I} = \frac{H}{\dot{I}} \l( 1 + \frac{\dot{H}}{H^2} \r) \approx - \frac{1}{M_{\rm P}^2} \frac{V}{V'} \l( 1 - \epsilon \r)
\eeq
Hence the running of slow-roll parameters are given by
\bea
\frac{d \epsilon}{d \ln k} &=& - 2 \epsilon \eta + 4 \epsilon^2
\\
\frac{d \eta}{d \ln k} &=& 2 \epsilon \eta - \xi^2
\\
\frac{d \xi^2}{d \ln k} &=& 4 \epsilon \xi^2 - \eta \xi^2 - \sigma^3
\eea
and, defining $\alpha_{\mathcal{R}} \equiv d n_s/d \ln k$ and $\alpha_T \equiv d n_T/d \ln k$, one finds
\bea 
\alpha_{\mathcal{R}} &=& - 8 \epsilon \l( 3 \epsilon - 2 \eta \r) - 2 \xi^2
\nonumber \\
&=& - \frac{1}{2} r \l( 1-n_s-\frac{3}{16}r \r) - 2 \xi^2
\\
\frac{d \alpha_{\mathcal{R}}}{d \ln k} &=& -32 \epsilon \l( \eta^2 - 6 \epsilon \eta + 6 \epsilon^2 \r) + 2 \l( \eta - 12 \epsilon \r) \xi^2 +2 \sigma^3
\nonumber \\
&=& - \frac{1}{2} r \l[ \l( 1 - n_s \r)^2 - \frac{3}{64} r^2 \r] - \l( 1 - n_s + \frac{9}{8} r \r) \xi^2 +2 \sigma^3
\\
\alpha_T &=& 4 \epsilon \l( \eta - 3 \epsilon \r)
\\
\frac{d \alpha_T}{d \ln k} &=& - 4 \epsilon \l( 2 \eta^2 - 18 \epsilon \eta + 24 \epsilon^2 + \xi^2 \r)
\nonumber \\
&=& - \frac{1}{4} r \l[ \frac{1}{2} \l( 1 - n_s \r)^2 + \frac{3}{16}r \l( 1 - n_s \r) - \frac{3}{64} r^2 + \xi^2 \r]
\eea

\section{Analytic expressions of slow-roll parameters in various models} 
\label{app:slow-roll-para}
In the following collections of formulas, $\phi_e$ stands for the field value at the end of inflation.
In cases of spiralized inflation models, we apply Eq.~(\ref{dVdI})-(\ref{d4VdI4}) instead of \eq{inflaton} in order not to miss relevant sub-leading terms, and ignore irrelevant higher order terms of $\gamma^2$ in expressions of slow-roll parameters. 

\subsection{Hilltop inflation (HI)}
The potential is 
\beq \label{RI}
V_\phi= V_0 \l[ 1 - \l( \frac{\phi}{\mu} \r)^p \r] + \dots
\eeq
where $V_0$ is the potential energy at $\phi=0$, $\mu$ is a mass parameter, $\dots$ denotes at least term(s) for stabilization, and we consider $p=4$ case only.
%One finds
%\bea
%&V_\phi' = - p \frac{V_0}{\mu^p} \phi^{p-1}, \quad V_\phi'' = -p (p-1) \frac{V_0}{\mu^p} \phi^{p-2}&
%\\
%&V_\phi''' = -p (p-1) (p-2) \frac{V_0}{\mu^p} \phi^{p-3}, \quad V_\phi'''' = -p (p-1) (p-2) (p-3) \frac{V_0}{\mu^p} \phi^{p-4}&
%\eea
Slow-roll parameters are
\bea
\epsilon &=& \frac{p^2}{2} \l( \frac{M_{\rm P}}{\mu} \r)^2 \l| \frac{\l( \phi/\mu \r)^{p-1}}{1 - (\phi/\mu)^p} \r|^2
\\
\eta &=& - p (p-1) \l( \frac{M_{\rm P}}{\mu} \r)^2 \frac{\l( \phi/\mu \r)^{p-2}}{1 - (\phi/\mu)^p},
\\
\xi^2 &=& p^2 (p-1) (p-2) \l( \frac{M_{\rm P}}{\mu} \r)^4 \frac{(\phi/\mu)^{2(p-2)}}{\l[ 1 - (\phi/\mu)^p \r]^2},
\\
\sigma^3 &=& - p^3 (p-1) (p-2) (p-3) \l( \frac{M_{\rm P}}{\mu} \r)^6 \frac{(\phi/\mu)^{3(p-2)}}{\l[ 1 - (\phi/\mu)^p \r]^3}
\eea
For the cosmologically relevant scales, $\phi \ll \mu$ and $\epsilon \ll \eta$ leading to  
\beq
n_s \simeq 1 + 2 \eta
\eeq 
%For $p=2$, $\phi_e$ satisfying $\epsilon=1$ is 
%\beq
%\phi_e =  \l[ 1 + \frac{1}{2} \l( \frac{M_{\rm P}}{\mu} \r)^2 \r]^{1/2} - \frac{1}{\sqrt{2}} \frac{M_{\rm P}}{\mu} 
%\eeq
For $p=4$, $\phi_e$ at $\epsilon=1$ can be found numerically.
The $e$-folding number for $p>2$ is
%\beq
%N_e^{\rm HI} \approx 
%\l\{
%\begin{array}{lc}
%\frac{\sqrt{2}}{2} \frac{\mu^2}{M_{\rm P}^2} \ln \frac{\phi_e}{\phi} & \textrm{for} \ p=2
%\\
%\frac{\sqrt{2}}{p (p-2)} \frac{\mu^2}{M_{\rm P}^2} \l[ \l( \frac{\mu}{\phi} \r)^{p-2} - \l( \frac{\mu}{\phi_e} \r)^{p-2} \r] & \textrm{for} \ p>2
%\end{array}
%\r.
%\eeq
\beq
N_e^{\rm HI} \approx 
\frac{\sqrt{2}}{p (p-2)} \frac{\mu^2}{M_{\rm P}^2} \l[ \l( \frac{\mu}{\phi} \r)^{p-2} - \l( \frac{\mu}{\phi_e} \r)^{p-2} \r] 
\eeq

\subsection{$R^2$-inflation (R2I)}
The potential is 
\beq \label{RI}
V_\phi= V_0 \l( 1 - e^{-\phi/\mu} \r)^2
\eeq
%It which gives
%\bea
%V_\phi' &=& \phantom{-} 2 V_0 c e^{-\phi/\mu} \l( 1 - e^{-\phi/\mu} \r)
%\\
%V_\phi'' &=& - 2 V_0 c^2 e^{-\phi/\mu} \l( 1 - 2 e^{-\phi/\mu} \r)
%\\
%V_\phi''' &=& \phantom{-} 2 V_0 c^3 e^{-\phi/\mu} \l( 1 - 4 e^{-\phi/\mu} \r)
%\\
%V_\phi'''' &=& - 2 V_0 c^4 e^{-\phi/\mu} \l( 1 - 8 e^{-\phi/\mu} \r)
%\eea
Slow-roll parameters are 
\bea
\epsilon = 2 \l( \frac{M_{\rm P}}{\mu} \r)^2 \frac{e^{-2\phi/\mu}}{\l( 1 - e^{-\phi/\mu} \r)^2} &,& \ \eta = - 2 \l( \frac{M_{\rm P}}{\mu} \r)^2 \frac{e^{-\phi/\mu} \l( 1 - 2 e^{-\phi/\mu} \r)}{\l( 1 - e^{-\phi/\mu} \r)^2}
\\
\xi^2 = 4 \l( \frac{M_{\rm P}}{\mu} \r)^4 \frac{e^{-2 \phi/\mu} \l( 1 - 4 e^{-\phi/\mu} \r)}{\l(1 - e^{-\phi/\mu}\r)^3} &,& \ \sigma^3 = -8 \l( \frac{M_{\rm P}}{\mu} \r)^6 \frac{e^{-3 \phi/\mu} \l( 1 - 8 e^{-\phi/\mu} \r)}{\l(1 - e^{-\phi/\mu}\r)^4}
\eea
Note that
\beq
\epsilon \simeq \frac{1}{2} \l( \frac{\mu}{M_{\rm P}} \r)^2 \eta^2
\eeq
Hence, taking a large $\mu$, one can get a larger $\epsilon$ realizing a large tensor-to-scalar ratio.
Taking $\mu = \sqrt{3/2} M_{\rm P}$ for the original $R^2$-inflation, one finds $\epsilon \simeq \frac{3}{4} \eta^2$ leading to 
\beq
n_s \simeq 1 + 2 \eta
\eeq 
From $\epsilon=1$, $\phi_e$ is given by 
\beq
\phi_e = \mu \ln \l[ 1 + \sqrt{2} (M_{\rm P}/\mu) \r]
\eeq
The $e$-folding number is
\beq
N_e^{\rm RI} = \frac{1}{2} \l(\frac{\mu}{M_{\rm P}} \r)^2 \l[ e^{\phi/\mu} - e^{\phi_e/\mu} - \frac{\l( \phi - \phi_e \r)}{\mu} \r]
\eeq

\subsection{Natural inflation (NI)}
The potential is
\beq \label{NI}
V = V_\phi = V_0 \l[ 1 + \cos(\phi/\mu) \r]
\eeq
%which gives
%\bea
%V_\phi' = - \frac{V_0}{M} \sin (\phi/M) &,& V_\phi'' = - \frac{V_0}{M^2} \cos (\phi/M)
%\\
%V_\phi''' = \frac{V_0}{M^3} \sin (\phi/M) &,& V_\phi'''' = \frac{V_0}{M^4} \cos (\phi/M)
%\eea
Slow-roll parameters are
\bea
\epsilon = \frac{1}{2} \l( \frac{M_{\rm P}}{\mu} \frac{\sin (\phi/M)}{\l[ 1 + \cos (\phi/\mu) \r]} \r)^2 &,& \ \eta = - \l( \frac{M_{\rm P}}{\mu} \r)^2 \frac{\cos (\phi/M)}{\l[ 1 + \cos (\phi/\mu) \r]}
\\
\xi^2 = - \l( \frac{M_{\rm P}}{\mu} \r)^4 \frac{1-\cos(\phi/\mu)}{1+\cos(\phi/\mu)} &,& \
\sigma^3 = \l( \frac{M_{\rm P}}{\mu} \r)^6 \frac{\cos (\phi/\mu) \l[ 1-\cos(\phi/\mu) \r]}{1+\cos(\phi/\mu)}
\eea
%\bea
%\xi^2 &=& - \l( \frac{M_{\rm P}}{M} \r)^4 \frac{1-\cos(\phi/M)}{1+\cos(\phi/M)}
%\\
%\sigma^3 &=& \l( \frac{M_{\rm P}}{M} \r)^6 \frac{\cos (\phi/M) \l[ 1-\cos(\phi/M) \r]}{1+\cos(\phi/M)}
%\eea
From $\epsilon=1$, $\phi_e$ is found to satisfy
\beq
\cos (\phi_e/\mu) \simeq - 1 + \l( \frac{M_{\rm P}}{\mu} \r)^2
\eeq
where $\mu \gg M_{\rm P}$ was assumed.
The $e$-folding number is 
\beq
N_e^{\rm NI} = \l( \frac{\mu}{M_{\rm P}} \r)^2 \ln \l[ \frac{1 - \cos (\phi_e/\mu)}{1 - \cos (\phi/\mu)} \r] \simeq \l( \frac{\mu}{M_{\rm P}} \r)^2 \ln \l[ \frac{2}{1 - \cos (\phi/\mu)} \r]
\eeq
%

%\subsection{Power-law potentials}
%The potential is given by a monomial of $\phi$ as
%\beq
%V = \lambda M_{\rm P}^4 \l( \frac{\phi}{M_{\rm P}} \r)^m
%\eeq
%giving 
%\bea
%\epsilon = \frac{n^2}{2} \l( \frac{M_{\rm P}}{\phi} \r)^2 &,& \eta = \frac{2 \l(n-1\r)}{n} \epsilon
%\\
%\xi^2 =  \frac{4 (n-1) (n-2)}{n^2} \epsilon^2 &,& \sigma^3 = \frac{8 (n-1) (n-2) (n-3)}{n^3} \epsilon^3
%\eea
%where $m$ is a positive fractional number.
%We consider only $m<2$.
%Note that $\xi^2=\sigma^3=0$ for $m=1$.

\subsection{Spiral chaotic inflation 1 (SCI1)}
The potential is
\beq
V = V_0 \l( \frac{\phi}{\mu} \r)^2 + V_{\rm M}
\eeq
Denoting a slow-roll parameter $x$ and $f(\phi)$ associated with a specific value of $n$ as $x_n$ and $f_n(\phi)$ respectively, one finds 
\bea
\epsilon_n &=& 2 \l( \gamma \frac{M_{\rm P}}{\phi} \r)^2
\\
\eta_n &=& \l[ - (n - 1) + n \gamma^2 \r] \epsilon_n
\\
\xi_n^2 &=& 2 n \l[ (n-1) - (3n-1) \gamma^2 \r] \epsilon_n^2
\\
\sigma_n^3 &=& \l\{ - 2n(n-1)(3n+1) + n \l[ (n-1)(36n+26)+24 \r] \gamma^2 \r\} \epsilon_n^3
\eea
From $\epsilon_n=1$, $\phi_e$ is given by
\beq
\phi_e = M \l( \frac{\sqrt{2}}{n} \frac{M_{\rm P}}{M} \r)^{\frac{1}{n+1}}
\eeq 
The $e$-folding number is 
\bea
N_{e,n}^{\rm SCI1} &=& \frac{1}{4} \frac{n^2}{n+1} \l( \frac{M}{M_{\rm P}} \r)^2 \l( \frac{\phi}{M} \r)^{2 (n+1)} \l[ 1 - \l( \frac{\phi_e}{\phi} \r)^{2(n+1)} \r] 
\nonumber \\
&=& \frac{f_n^2}{4 \l( n+1 \r)} \l( \frac{\phi}{M_{\rm P}} \r)^2 \l[ 1 - \l( \frac{\phi_e}{\phi} \r)^{2(n+1)} \r] 
\nonumber \\
&\approx& \frac{1}{2 \l( n+1 \r) \epsilon_n}
\eea
leading to
\beq
1-n_s \approx \frac{n+2}{n+1}\frac{1}{N_e} = \frac{n+2}{8} r
\eeq

\subsection{Spiral chaotic inflation 2 (SCI2)}
The potential is
\beq
V = V_0 \l( \frac{\phi}{\mu} \r)^4 + V_{\rm M}
\eeq
Slow-roll parameters are
\bea
\epsilon_n &=& 8 \l( \gamma \frac{M_{\rm P}}{\phi} \r)^2
\\
\eta_n &=& \frac{3-n}{2} \epsilon_n
\\
\xi_n^2 &=& \frac{1}{2} \l[ -(n-1)(3-n) + n (4-3n) \gamma^2 \r] \epsilon_n^2
\\
\sigma_n^3 &=& \frac{1}{4} \l\{ (n-1)(3-n)(3n-1) + n \l[ (n-1)(18n-17)-4 \r] \gamma^2 \r\}\epsilon_n^3
\eea
From $\epsilon=1$, $\phi_e$ is given by
\beq
\phi_e = M \l( \frac{2 \sqrt{2}}{n} \frac{M_{\rm P}}{M} \r)^{\frac{1}{n+1}}
\eeq
The $e$-folding number is 
\beq
N_{e,n}^{\rm SCI2} = \frac{f_n^2}{8 \l(n+1 \r)} \l( \frac{\phi}{M_{\rm P}} \r)^2 \l[ 1 - \l( \frac{\phi_e}{\phi} \r)^{2(n+1)} \r] \approx \frac{1}{(n+1) \epsilon_n}
\eeq
leading to
\beq
1-n_s \approx \frac{n+3}{n+1} \frac{1}{N_e} = \frac{n+3}{16} r
\eeq

\subsection{Spiral inflation (SI)}
The potential is
\beq
V = V_0 \l[ 1 - \l( \frac{\phi}{\phi_0} \r)^2 \r]^2 + V_{\rm M}
\eeq
Slow-roll parameters are 
\bea \label{SI-ep}
\epsilon_n &=& 8 \l( \gamma \frac{M_{\rm P} \phi}{\phi_0^2-\phi^2} \r)^2
\\ \label{SI-eta}
\eta_n &=& \frac{\epsilon_n}{2} \l\{ (n-1) \l( \frac{\phi_0}{\phi} \r)^2 + (3-n) - n \gamma^2 \l[ \l( \frac{\phi_0}{\phi} \r)^2-1 \r] \r\}
\\ \label{SI-xi2}
\xi_n^2 &=& \frac{\epsilon_n^2}{2} \l[ \l( \frac{\phi_0}{\phi} \r)^2 - 1 \r] \l\{ (n-1) \l[ n \l( \frac{\phi_0}{\phi} \r)^2 + (3-n) \r] - n \gamma^2 \l[ (3n-1) \l( \frac{\phi_0}{\phi} \r)^2 - (3n-4) \r] \r\}
\\ \label{SI-sigma3}
\sigma_n^3 &=& \frac{\epsilon_n^3}{4} \l[ \l( \frac{\phi_0}{\phi} \r)^2 - 1 \r]^2 \l\{ (n-1) \l[ n (3n+1) \l( \frac{\phi_0}{\phi} \r)^2 + (3-n)(3n-1) \r] \r.
\nonumber \\
&& \l. - n \gamma^2 \l[ \l( (n-1)(18n+13)+12 \r) \l( \frac{\phi_0}{\phi} \r)^2 - \l( (n-1)(18n-17)-4 \r) \r] \r\}
\eea
The $e$-folding number is given by
\beq \label{SI-Ne}
N_{e,n}^{\rm SI} = \frac{1}{8 n} \l( \frac{\phi_0}{M_{\rm P}} \r)^2 f_n^2(\phi_e) \l\{ \l[ 1 - \l( \frac{\phi}{\phi_e} \r)^{2n} \r] - \frac{n}{n+1} \l( \frac{\phi_e}{\phi_0} \r)^2 \l[ 1 - \l( \frac{\phi}{\phi_e} \r)^{2(n+1)} \r]  \r\}
\eeq
%
%The $e$-foldings are
%\beq \label{SI-Ne}
%N_e^{\rm SI} = \frac{1}{8 n} \l( \frac{\phi_0}{M_{\rm P}} \r)^2 \l\{ \l[ 1 - \frac{n}{n+1} \l( \frac{\phi_e}{\phi_0} \r)^2 \r] a_f^2 -  \l[ 1 - \frac{n}{n+1} \l( \frac{\phi_i}{\phi_0} \r)^2 \r] a_i^2 \r\}
%\eeq
%where $a_{i,f} = a(\phi_{i,f})$
%
If $\Lambda$ is small enough, $\phi_e$ can be from 
\beq
\l| \frac{\partial V_\phi}{\partial \phi} \r| \simeq \frac{f_n}{\phi} \Lambda^4
\eeq
equivalent to
\beq \label{SI-phi-f-cond}
\frac{4}{n} \l( \frac{M}{\phi_0} \r)^n \l( \frac{\phi}{\phi_0} \r)^{2-n} \l[ 1 - \l( \frac{\phi}{\phi_0} \r)^2 \r] \simeq \frac{\Lambda^4}{V_\phi(\phi=0)}
\eeq 
Otherwise, it is from $\epsilon_n=1$.
For $\phi_0 = M_{\rm P}$ which we assume for simplicity, if $M \ll \phi_0$ which is true in spiraling inflation models of our consideration, $\epsilon_n=1$ gives solutions $\phi_e \ll \phi_0$ or $\phi_e \sim \phi_0$.
The former is not a proper solution for $\phi > M$, and the latter is  
\beq
\l. \frac{\phi_e}{\phi_0} \r|_{\epsilon=1} \approx \l[ 1 - \frac{2 \sqrt{2}}{n} \frac{M^n M_{\rm P}}{\phi_0^{n+1}} \r]^{1/2}
\eeq

For $n=1$, defining $\kappa \equiv M \phi_0 / \sqrt{2} \phi^2$ for convenience, one finds
\beq \label{SI-eta1}
\eta_1 \simeq \epsilon_1 \l( 1 - \kappa^2 \r) \ \Rightarrow \ \epsilon_1 = \frac{1-n_s}{6-2(1-\kappa^2)}
\eeq 
Note that $\eta_1$ can be either positive or negative.
If $\kappa \ll 1$, $\epsilon_1 \approx \eta_1$ which results in 
\bea
n_s &=& 1- \frac{r}{4} \gtrsim 0.975
\\ \label{SI-M1}
\frac{M}{\phi_0} &\simeq& 0.028 \l[ 1 - \l( \frac{\phi_*}{\phi_0} \r)^2 \r] \l( \frac{\phi_0}{M_{\rm P}} \r) \l( \frac{r}{0.1} \r)^{1/2} \ll \sqrt{2} \l(\frac{\phi}{\phi_0} \r)^2
\eea
where the lower bound of $n_s$ is due to $r< 0.1$ from observations \cite{Ade:2015lrj}, and we assumed $\phi^2 \ll \phi_0$ which is true for cosmological scales relevant CMB observations.
If inflation ends at $\epsilon_1=1$, from \eqs{SI-Ne}{SI-M1} with $\phi_e \simeq \phi_0$ the number of $e$-foldings is found to be 
\beq
N_{e, \epsilon=1}^{\rm SI} \simeq 80 \l( \frac{0.1}{r} \r)
\eeq
which does not depend on $\phi_*$ and too large to match observation unless $r \gtrsim 0.14$.
Hence, in order to match observations, inflation should end by waterfall drop at $\phi_e < \phi_\times = \phi_0/\sqrt{3}$ with $\phi_\times$ being the maximum field value satisfying \eq{SI-phi-f-cond}.
In such a case, the $e$-foldings can be around $40$ at most which is too small to match observations.
However, note that the left-hand side of \eq{SI-phi-f-cond} decreases for $\phi > \phi_\times$, allowing the possibility of a two-step inflation (before and after $\phi=\phi_\times$) in which inflation eventually ends when $\epsilon_1=1$.
Adjusting $\Lambda$ which does not affect slow-roll parameters, one can control $\phi_e$ via \eq{SI-phi-f-cond} to reduce the total $e$-foldings.  
Hence, even if $N_e^{\rm SI}$ for $\phi < \phi_\times$ is too small to match observations by itself, it is not a problem as long as it covers observed CMB scales and $e$-foldings for $\phi>\phi_\times$ are large enough.
Moreover, the required $e$-foldings for primordial inflation can be reduced in cases of long period of matter domination after inflation, a very low reheating temperature close to its lower bound, or some extra $e$-foldings, for example, from thermal inflation \cite{Lyth:1995hj,Lyth:1995ka}.
Again, the need of all these possibility depend on $\Lambda$ which does not affect slow-roll parameters.  
In this work, we do not pursue the details of these possibilites.

On the other hand, if $\kappa \gg 1$, $\eta_1$ becomes negative and $r$ can be well below observational bound even for the preferred cental value of $n_s$.
%the first two slow-roll parameters become
%\bea
%\epsilon_1 &\simeq& 16 \kappa^2 \l[ \l( \frac{\phi}{\phi_0} \r)^2 \frac{M_{\rm P}}{\phi_0} \r]^2 \l[ 1 - \l( \frac{\phi}{\phi_0} \r)^2 \r]^{-2} 
%\\
%\eta_1 &\simeq& - \kappa^2 \epsilon_1
%\eea
%leading to 
%\beq
%\epsilon_1 \simeq \frac{1-n_s}{6+2\kappa^2}
%\eeq
Note that 
$\phi_*$ is constrained as 
\beq
\frac{M}{\phi_0} \lesssim \frac{\phi_*}{\phi_0} \ll \l( \frac{1}{\sqrt{2}} \frac{M}{\phi_0} \r)^{1/2}
\eeq
Hence, $M/\phi_0$ should be smaller than unity by at least a couple of orders of magnitude in order to allow an enough room for $\phi_*$. 
In this case, the spectral running and the running of the running become large, allowing easy discrimination in the future experiments. 
%again the $e$-foldings matching observation can be obtained by adjusting waterfall end of inflation. 

For $n=2$, \eq{SI-phi-f-cond} has a solution at  
\beq
\l. \frac{\phi_\times}{\phi_0} \r|_{\epsilon_2<1} = 1 - \frac{1}{2} \l( \frac{\phi_0}{M} \r)^2 \frac{\Lambda^4}{V_\phi(\phi=0)}
\eeq
%which can be made small by taking a small $\Lambda$.
Contrary to $n=1$ case, the field configuration can follow the spiraling trench only for $\phi>\phi_\times$ and never gets out, and inflation ends only when $\epsilon_2=1$ satisfied at
\beq
\l. \frac{\phi_e}{\phi_0} \r|_{\epsilon_2=1} \approx \l[ 1 - \l( \frac{\sqrt{2} M^2 M_{\rm P}}{\phi_0^3} \r) \r]^{1/2}
\eeq
for $M \ll \phi_0$.
Meanwhile, from \eq{SI-ep} one find that 
\beq \label{SI-M2}
\l( \frac{M_{\rm P}}{\phi_0} \r)^{1/2}\frac{M}{\phi_0} = \l( \frac{r}{32} \r)^{1/4} \l[ \l( \frac{\phi_*}{\phi_0} \r) \l( 1- \l( \frac{\phi_*}{\phi_0} \r)^2 \r) \r]^{1/2}   \lesssim 0.1467 \l( \frac{r}{0.1} \r)^{1/4}
\eeq 
which means $\phi_e \simeq \phi_0$.
Adjusting $\Lambda$, one can make $\phi_\times < \phi_*$.
Then, combined with \eq{SI-M2}, the number of $e$-folding with $\phi_e \approx \phi_0$ is minimized at $\phi_*/\phi_0 \simeq 0.6356$ with  
\beq
N_{e,n=2}^{\rm SI} \simeq 5.72 \times \frac{16}{3 r} \simeq 305 \l( \frac{0.1}{r} \r)
\eeq
which is too large to match observations, and we do not consider this case any longer in regard of the spectral running and its running.

\subsection{Spiral Coleman-Weinberg inflation (SCWI)}
The potential is
\beq
V = V_0 \l\{ 1 + 4  \l( \frac{\phi}{\phi_0} \r)^4 \l[ \ln \l( \frac{\phi}{\phi_0} \r) - \frac{1}{4} \r]  \r\} + V_{\rm M}
\eeq
Slow-roll parameters are
\bea
\epsilon_n &=& 128 \l[ \gamma \frac{M_{\rm P} \phi^3}{\phi_0^4} \ln \l( \frac{\phi}{\phi_0} \r) \r]^2 \l( \frac{V_0}{V} \r)^2
\\
\eta_n  
&=& \frac{1}{8} \l( \frac{\phi_0}{\phi} \r)^4 \l( \frac{V}{V_0} \r) \frac{\l\{ \l[3-n(1-\gamma^2) \r] \ln \l( \frac{\phi}{\phi_0} \r) + 1 \r\}}{\ln^2 \l( \frac{\phi}{\phi_0} \r)} \epsilon_n
\\ \label{SCWI-xi2}
\xi_n^2 &=& \frac{\l\{ b_n \ln \l( \frac{\phi}{\phi_0} \r) + \l[5-3n(1-\gamma^2)\r] \r\} \ln \l( \frac{\phi}{\phi_0} \r)}{\l\{ \l[ 3-n(1-\gamma^2) \r] \ln \l( \frac{\phi}{\phi_0} \r) +1 \r\}^2} \eta_n^2
\\ \label{SCWI-sigma3}
\sigma_n^3 &=& \frac{\l[ c_n \ln \l( \frac{\phi}{\phi_0} \r) + d_n \r]}{8 \l\{ \l[ 3-n(1-\gamma^2) \r] \ln \l( \frac{\phi}{\phi_0} \r) +1 \r\}^2} \l( \frac{\phi_0}{\phi} \r)^4 \l( \frac{V}{V_0} \r) \epsilon_n \eta_n^2
\eea
where
\bea
b_n &\equiv& -2 (n-1)(3-n) + 2 \gamma^2 n(4-3n)
\\
c_n &\equiv& 2 (n-1)(3-n)(3n-1) + 2 \gamma^2 n(2n-1)(9n-13)
\\
d_n &\equiv& (n-1)(11n-15)-4 + 2 \gamma^2 n(13-15n)
\eea
Depending on the maginitude of $\Lambda$ relative to $V_0^{1/4}$, inflation can end by waterfall drap at $\phi_e$ satisfying 
\beq
\l| \frac{\partial V_\phi}{\partial \phi} \r| \simeq \frac{f_n}{\phi} \Lambda^4
\eeq
equivalent to
\beq \label{SCW-phix}
\frac{16}{n} \l( \frac{M}{\phi_0} \r)^n \l( \frac{\phi_e}{\phi_0} \r)^{4-n} \ln \l( \frac{\phi_0}{\phi_e} \r) \simeq \frac{\Lambda^4}{V_\phi(\phi=0)}
\eeq 
The left-hand side of the equation above is maximized at $\phi_\times/\phi_0=e^{-\frac{1}{4-n}}$.
%
%For $n=1$, if $\epsilon_1 \ll \eta_1$ that will be justified shortly, 
%\beq
%\l( \frac{M M_{\rm P}}{\phi_0^2} \r)^2 \simeq \frac{n_s-1}{32 \l[ \ln (\phi/\phi_0) + 1\r] (V_0/V(\phi))}
%\eeq
%leading to 
%\beq
%\epsilon_1 \approx \frac{4 (n_s-1) \ln^2 (\phi/\phi_0)}{2 \ln (\phi/\phi_0) +1} \l( \frac{V_0}{V(\phi)} \r) \l( \frac{\phi}{\phi_0} \r)^4
%\eeq
%If $\phi_e < \phi_\times$, the number of $e$-foldings is bounded 
%\bea
%N_e^{\rm SICW} 
%&\lesssim& \frac{1}{16} \l( \frac{\phi_0^2}{M M_{\rm P}} \r)^2 \ln \l[ \frac{\ln (\phi/\phi_0)}{\ln (\phi_e/\phi_0)} \r]
%\nonumber \\
%&\approx& \frac{2 \l[ 2 \ln (\phi/\phi_0) +1 \r]}{n_s - 1} \l( \frac{V_0}{V(\phi)} \r) \ln \l[ \frac{\ln (\phi/\phi_0)}{\ln (\phi_e/\phi_0)} \r]
%\eea
If a solution to \eq{SCW-phix} is absent, inflation ends when $\epsilon_n=1$ at $\phi_e$ satisfying
\beq
\frac{V_0}{V(\phi_e)} \l( \frac{\phi_e}{\phi_0} \r)^{3-n} \ln \l( \frac{\phi_0}{\phi_e} \r) = \frac{1}{8 \sqrt{2}} \l( \frac{\phi_0^{n+1}}{M^n M_{\rm P}} \r)
\eeq
Formally, the number of $e$-foldings is given by
\beq
N_{e,n}^{\rm SCWI} = \frac{1}{M_{\rm P}} \int \frac{dI}{\sqrt{2 \epsilon_n}} \approx \frac{1}{M_{\rm P}} \int d \phi \frac{f_n}{c_\theta \sqrt{2 \epsilon_n}}
\eeq
but it can not be given as a simple closed analytic form.

\acknowledgments

The authors acknowledge support from the MEC and FEDER (EC) Grants FPA2011-23596 and the Generalitat Valenciana under grant PROMETEOII/2013/017.
G.B. acknowledges partial support from the European Union FP7 ITN INVISIBLES (Marie Curie Actions, PITN-GA-2011-289442).

%\paragraph{Note added.} This is also a good position for notes added after the paper has been written.



\begin{thebibliography}{99}

%\cite{Guth:1980zm}
\bibitem{Guth:1980zm} 
  A.~H.~Guth,
  %``The Inflationary Universe: A Possible Solution to the Horizon and Flatness Problems,''
  Phys.\ Rev.\ D {\bf 23}, 347 (1981).


%\cite{Guth:1979bh}
\bibitem{Guth:1979bh} 
  A.~H.~Guth and S.~H.~H.~Tye,
  %``Phase Transitions and Magnetic Monopole Production in the Very Early Universe,''
  Phys.\ Rev.\ Lett.\  {\bf 44}, 631 (1980)
  [Erratum-ibid.\  {\bf 44}, 963 (1980)].


%\cite{Mukhanov:1981xt}
\bibitem{Mukhanov:1981xt} 
  V.~F.~Mukhanov and G.~V.~Chibisov,
  %``Quantum Fluctuation and Nonsingular Universe. (In Russian),''
  JETP Lett.\  {\bf 33}, 532 (1981)
  [Pisma Zh.\ Eksp.\ Teor.\ Fiz.\  {\bf 33}, 549 (1981)].

%\cite{Mukhanov:1982nu}
\bibitem{Mukhanov:1982nu} 
  V.~F.~Mukhanov and G.~V.~Chibisov,
  %``The Vacuum energy and large scale structure of the universe,''
  Sov.\ Phys.\ JETP {\bf 56}, 258 (1982)
  [Zh.\ Eksp.\ Teor.\ Fiz.\  {\bf 83}, 475 (1982)].



%\cite{Silverstein:2008sg}
\bibitem{Silverstein:2008sg} 
  E.~Silverstein and A.~Westphal,
  %``Monodromy in the CMB: Gravity Waves and String Inflation,''
  Phys.\ Rev.\ D {\bf 78}, 106003 (2008)
  [arXiv:0803.3085 [hep-th]].

%\cite{McAllister:2008hb}
\bibitem{McAllister:2008hb} 
  L.~McAllister, E.~Silverstein and A.~Westphal,
  %``Gravity Waves and Linear Inflation from Axion Monodromy,''
  Phys.\ Rev.\ D {\bf 82}, 046003 (2010)
  [arXiv:0808.0706 [hep-th]].
  
%\cite{Berg:2009tg}
\bibitem{Berg:2009tg} 
  M.~Berg, E.~Pajer and S.~Sjors,
  %``Dante's Inferno,''
  Phys.\ Rev.\ D {\bf 81}, 103535 (2010)
  [arXiv:0912.1341 [hep-th]].  

%\cite{McDonald:2014oza}
\bibitem{McDonald:2014oza} 
  J.~McDonald,
  %``Sub-Planckian Two-Field Inflation Consistent with the Lyth Bound,''
  JCAP {\bf 1409}, no. 09, 027 (2014)
  [arXiv:1404.4620 [hep-ph]].

%\cite{McDonald:2014nqa}
\bibitem{McDonald:2014nqa} 
  J.~McDonald,
  %``A Minimal Sub-Planckian Axion Inflation Model with Large Tensor-to-Scalar Ratio,''
  arXiv:1407.7471 [hep-ph].

%\cite{Li:2014vpa}
\bibitem{Li:2014vpa} 
  T.~Li, Z.~Li and D.~V.~Nanopoulos,
  %``Helical Phase Inflation,''
  arXiv:1409.3267 [hep-th].

%\cite{Carone:2014cta}
\bibitem{Carone:2014cta} 
  C.~D.~Carone, J.~Erlich, A.~Sensharma and Z.~Wang,
  %``Dante's Waterfall,''
  arXiv:1410.2593 [hep-ph].

%\cite{Barenboim:2014vea}
\bibitem{Barenboim:2014vea} 
  G.~Barenboim and W.~I.~Park,
  %``Spiral Inflation,''
  Phys.\ Lett.\ B {\bf 741}, 252 (2015)
  [arXiv:1412.2724 [hep-ph]].

%\cite{Barenboim:2015zka}
\bibitem{Barenboim:2015zka} 
  G.~Barenboim and W.~I.~Park,
  %``Spiral Inflation with Coleman-Weinberg Potential,''
  Phys.\ Rev.\ D {\bf 91}, no. 6, 063511 (2015)
  [arXiv:1501.00484 [hep-ph]].

%\cite{Li:2015mwa}
\bibitem{Li:2015mwa} 
  T.~Li, Z.~Li and D.~V.~Nanopoulos,
  %``Symmetry Breaking Indication for Supergravity Inflation in Light of the Planck 2015,''
  arXiv:1502.05005 [hep-ph].

%\cite{Boubekeur:2005zm}
\bibitem{Boubekeur:2005zm} 
  L.~Boubekeur and D.~H.~Lyth,
  %``Hilltop inflation,''
  JCAP {\bf 0507}, 010 (2005)
  [hep-ph/0502047].

%\cite{Starobinsky:1980te}
\bibitem{Starobinsky:1980te} 
  A.~A.~Starobinsky,
  %``A New Type of Isotropic Cosmological Models Without Singularity,''
  Phys.\ Lett.\ B {\bf 91}, 99 (1980).

%\cite{Freese:1990rb}
\bibitem{Freese:1990rb}
  K.~Freese, J.~A.~Frieman and A.~V.~Olinto,
  %``Natural inflation with pseudo - Nambu-Goldstone bosons,''
  Phys.\ Rev.\ Lett.\  {\bf 65} (1990) 3233.


%\cite{Ade:2015lrj}
\bibitem{Ade:2015lrj} 
  P.~A.~R.~Ade {\it et al.} [Planck Collaboration],
  %``Planck 2015 results. XX. Constraints on inflation,''
  arXiv:1502.02114 [astro-ph.CO].
  
%\cite{Linde:1983gd}
\bibitem{Linde:1983gd} 
  A.~D.~Linde,
  %``Chaotic Inflation,''
  Phys.\ Lett.\ B {\bf 129}, 177 (1983).

%\cite{Lyth:1995hj}
\bibitem{Lyth:1995hj} 
  D.~H.~Lyth and E.~D.~Stewart,
  %``Cosmology with a TeV mass GUT Higgs,''
  Phys.\ Rev.\ Lett.\  {\bf 75}, 201 (1995)
  [hep-ph/9502417].

%\cite{Lyth:1995ka}
\bibitem{Lyth:1995ka} 
  D.~H.~Lyth and E.~D.~Stewart,
  %``Thermal inflation and the moduli problem,''
  Phys.\ Rev.\ D {\bf 53}, 1784 (1996)
  [hep-ph/9510204].

\bibitem{EUCLID}
  L.~Amendola {\it et al.} [Euclid Theory Working Group Collaboration],
  %``Cosmology and fundamental physics with the Euclid satellite,''
  Living Rev.\ Rel.\  {\bf 16}, 6 (2013)
  [arXiv:1206.1225 [astro-ph.CO]].

\bibitem{SPHEREx}
  O.~Doré {\it et al.},
  %``Cosmology with the SPHEREX All-Sky Spectral Survey,''
  arXiv:1412.4872 [astro-ph.CO].
  
%\cite{Peterson:2010mv}
\bibitem{Peterson:2010mv} 
  C.~M.~Peterson and M.~Tegmark,
  %``Non-Gaussianity in Two-Field Inflation,''
  Phys.\ Rev.\ D {\bf 84}, 023520 (2011)
  doi:10.1103/PhysRevD.84.023520


% Please avoid comments such as "For a review'', "For some examples",
% "and references therein" or move them in the text. In general,
% please leave only references in the bibliography and move all
% accessory text in footnotes.

% Also, please have only one work for each \bibitem.


\end{thebibliography}
\end{document}